\documentclass[apjl]{emulateapj} \usepackage{apjfonts} \newcommand{\markcite}[1]{}

\newcommand{\figref}[1]{Figure \ref{#1}}
\newcommand{\eqref}[1]{equation (\ref{#1})}

\newcommand{\sub}[1]{\ensuremath{_{\mbox{\scriptsize#1}}}}

\begin{document}
\slugcomment{accepted for publication in the Astrophysical Journal}
\title{Signatures of Planet Formation in Gravitationally Unstable Disks}
\shorttitle{SIGNATURES OF PLANET FORMATION}
\author{Hannah Jang-Condell \& Alan P.~Boss}
\shortauthors{JANG-CONDELL \& BOSS}
\affil{Carnegie Institution of Washington, Department of Terrestrial Magnetism}
\affil{5241 Broad Branch Road NW, Washington, D.C. 20015}

\begin{abstract}
In this paper, we calculate simulated scattered light 
images of a circumstellar disk in which a planet is forming by 
gravitational instability.  The simulated images bear no correlation 
to the vertically integrated surface density of the disk, but 
rather trace the density structure in the tenuous upper layers of the disk.  
Although the density at high altitudes does not bear a direct relation 
to activity at the midplane, the very existence of structure 
at high altitudes along with high time variability is an indicator 
of gravitational instability within the disk.  The timescale for variations 
is much shorter than the orbital period of the planet, which 
facilitates observation of the phenomenon.  Scattered light images 
may not necessarily be able to tell us where exactly a planet might 
be forming in a disk, but can still be a useful probe of 
active planet formation within a circumstellar disk.  
Although these phenomena are unlikely to be observable 
by current telescopes, future large telescopes, such as the 
Giant Magellan Telescope, may be able to detect them.  
\end{abstract}


\section{Introduction}

There is an ongoing theoretical debate about the dominant
formation mechanism for giant planets. Two opposite extremes
exist: core accretion and disk instability. In core accretion,
a solid core forms first by the collisional accumulation of
planetesimals and planetary embryos, and then accretes a
gaseous envelope from the protoplanetary disk gas \markcite{1980Mizuno}({Mizuno} 1980).
In the disk instability mechanism, a gaseous protoplanet
forms first in a gravitationally unstable, gaseous disk \markcite{boss97}({Boss} 1997),
and then a solid core forms by the settling and coagulation of
dust grains within the protoplanet. While core accretion is
generally assumed to be the dominant formation mechanism for
giant planets, considerable theoretical uncertainty remains about
both formation mechanisms \markcite{2007Durisen_etal}(e.g., {Durisen} {et~al.} 2007). Given this
murky situation, observational constraints are likely to be crucial
for deciding which mechanism, if either, dominates \markcite{2006Boss}({Boss} 2006).

Millimeter wave telescopes should be able probe the
planet-forming midplanes of optically-thick disks. The Atacama
Large Millimeter Array (ALMA) may be able to directly image
gaseous protoplanets formed by disk instability, or to detect the
strong self-absorption of HCO$^+$ emission lines caused by
these dense clumps of molecular gas \markcite{2006Narayanan}({Narayanan} {et~al.} 2006).
The latter type of detections might also be achieved by the
Submillimeter Array (SMA).

Here we consider the observational appearance of light scattered
by the irregular surfaces of a marginally gravitationally unstable
disk that is in the process of forming gaseous protoplanets
by the disk instability mechanism. The disk model studied here
\markcite{boss01}(from {Boss} 2001) is the same as the one investigated by 
\markcite{2006Narayanan}{Narayanan} {et~al.} (2006), and this work thus provides an important comparision
between what can be accomplished at optical and at millimeter
wavelengths in terms of constraining giant planet formation
mechanisms.

Many others have studied the observable properties of 
YSOs (young stellar objects).
This includes the spectral energy distributions 
of unresolved sources 
\markcite{1993ApJ...411..274Y,1995ApJ...439L..55B,1996ApJ...469..366B,1998ApJ...497..404W,2002ApJ...567.1183W,2006Robitaille_etal}(e.g.~ {Yorke}, {Bodenheimer}, \&  {Laughlin} 1993; {Boss} \& {Yorke} 1995, 1996; {Wood} {et~al.} 1998, 2002; {Robitaille} {et~al.} 2006)
and imaging of smooth axisymmetric disk models 
\markcite{1992ApJ...395..529W,1993ApJ...402..605W,1993ApJ...411..274Y,1995A&A...299..545S,1998A&A...337..832K,1999ApJ...525..330Y,2004Lucas_etal}(e.g.~ {Whitney} \& {Hartmann} 1992, 1993; {Yorke} {et~al.} 1993; {Sonnhalter}, {Preibisch}, \&  {Yorke} 1995; {Kessel}, {Yorke}, \&  {Richling} 1998; {Yorke} \& {Bodenheimer} 1999; {Lucas} {et~al.} 2004).  
In this paper, we discuss high-resolution imaging of a disk 
undergoing a dynamical process.  The basis for our modeling is a 
three-dimensional hydrodynamic simulation rather than an analytic disk 
model since we are interested in 
the perturbations that indicate activity in the disk rather than 
the overall emission. 

\section{Model Calculation}

\subsection{Disk Structure}

The disk model we adopt was calculated by \markcite{boss01}{Boss} (2001), and is described 
in detail in that paper.  
The hydrodynamics of the disk were calculated three-dimensionally in 
spherical coordinates, with $r$ as the distance from the central star, 
$\theta$ as the angle with respect to the axis of the disk's rotation, and 
$\phi$ as the azimuthal or longitudinal angle.  
Pressures and temperatures were calculated on a grid 
with $N_r=100$ uniformly spaced intervals in $r$ between 4 and 20 AU 
and $N_{\phi}=512$ uniformly spaced intervals in $\phi$ 
between 0 and $2\pi$.  In $\theta$, $N_{\theta}=23$ values were 
sampled, with $0\degr\leq\theta\leq90\degr$.  
The system consists of a 1 $M_{\sun}$ central protostar, 
surrounded by a gaseous disk of mass $0.091 M_{\sun}$ between 4 and 20 AU.
From an initially smooth disk, the system was 
evolved several hundred years, until a gravitationally bound 
clump began to form.  
In this study, we will examine snapshots of the simulation 
at 335, 339, and 346 years.

\subsection{Surface}

Simulated scattered light images are generated by treating the 
surface of the disk as an isotropic scatterer, with half 
the incident flux being absorbed and remainder scattered isotropically.  
The surface of the disk is defined to be where the 
disk becomes optically thick to stellar irradiation.
That is, if we define $\tau_s$ to be the optical depth, 
the surface is where $\tau_s=2/3$.  We assume the dust 
is well-mixed with the gas, using opacities 
calculated by \markcite{dalessio3}{D'Alessio}, {Calvet}, \&  {Hartmann} (2001) for a grain size distribution 
with maximum grain size $a\sub{max} = 1$ mm, power-law index of 
$p = 3.5$, with a dust temperature of 100K.  
They calculate the Planck opacity 
integrated over a stellar blackbody spectrum at 4000 K as
$\kappa^*_P
= 11.7$ g$^{-1}$ cm$^{2}$.  
The peak wavelength for this spectrum is 0.73 $\mu$m,
corresponding to R-band.  
We treat the star as a blackbody of radius $R_* = 2.6 R_{\sun}$ 
radiating at a temperature of $T_* = 4000$ K.

The surface is defined to be where the line-of-sight optical 
depth to the star is $\tau_s= 2/3$.  This is done numerically by 
calculating along lines of sight from the star. 
To avoid numerical artifacts at the inner boundary, we skip the innermost 
5 radial grid values, so that $r\sub{in}=r_5.$
We assume that the density interior to 
$r\sub{in}$ is constant ($\rho\sub{in}=\rho(r\sub{in})$), so 
\begin{equation}
\tau_s(r,\theta,\phi) = \left\{\begin{array}{ll}
\kappa^*_P\, \rho\sub{in}(\theta,\phi)\, r, & r<r\sub{in} \\
\kappa^*_P\, \rho\sub{in}(\theta,\phi)\, r\sub{in} + 
\int_{r\sub{in}}^r \kappa^*_P\, \rho(r',\theta,\phi) \;dr', &
r>r\sub{in}
\end{array}\right.
\end{equation}
For a line-of-sight along a given direction $(\theta,\phi)$, 
we calculate where the optical depth reaches 2/3 and define that to be 
the location of the disk surface, sampling it with a resolution of 
$\Delta r/r < \epsilon$ or 
$\Delta \theta < \epsilon \mu\sub{min}/2 $, 
where 
\(\mu\sub{min} = 4R_*/(3\pi r)
\)
and 
$\epsilon \approx [\ln(r\sub{max})-\ln(r\sub{min})]/N_r = 0.016$.

\subsection{Scattered flux}

The flux of stellar irradiation incident on the disk at 
a distance $r$ is 
\begin{equation}
F\sub{inc} = \sigma_B T_{\star}^4 \left(\frac{R_{\star}}{r}\right)^2.
\end{equation}
The angle of incidence of stellar irradiation on the disk surface is 
\begin{equation}
\mu = \mathbf{\hat{r}} \cdot \mathbf{\hat{n}} + \mu\sub{min}
\end{equation}
where $\mathbf{\hat{r}}$ points from the surface to the star, 
and $\mathbf{\hat{n}}$ is normal to the surface.  
The finite size of the star is the reason for including 
the $\mu\sub{min}$ term.  
To calculate the scattered flux, we 
assume that surface acts like an isotropic scatterer so that 
half the flux is absorbed into the disk and reprocessed into thermal 
radiation and the other half is scattered back out 
isotropically.  If we define $F_0=\sigma_B T_{\star}^4 R_{\star}^2$
to be the flux emitted by the star, then we can express the 
scattered light emission relative to the stellar flux:
\begin{equation}
F\sub{sca} = \frac{\mu F_0}{2 r^2}.
\end{equation}

\section{Results}

\begin{figure*}
\resizebox{\textwidth}{!}{\includegraphics{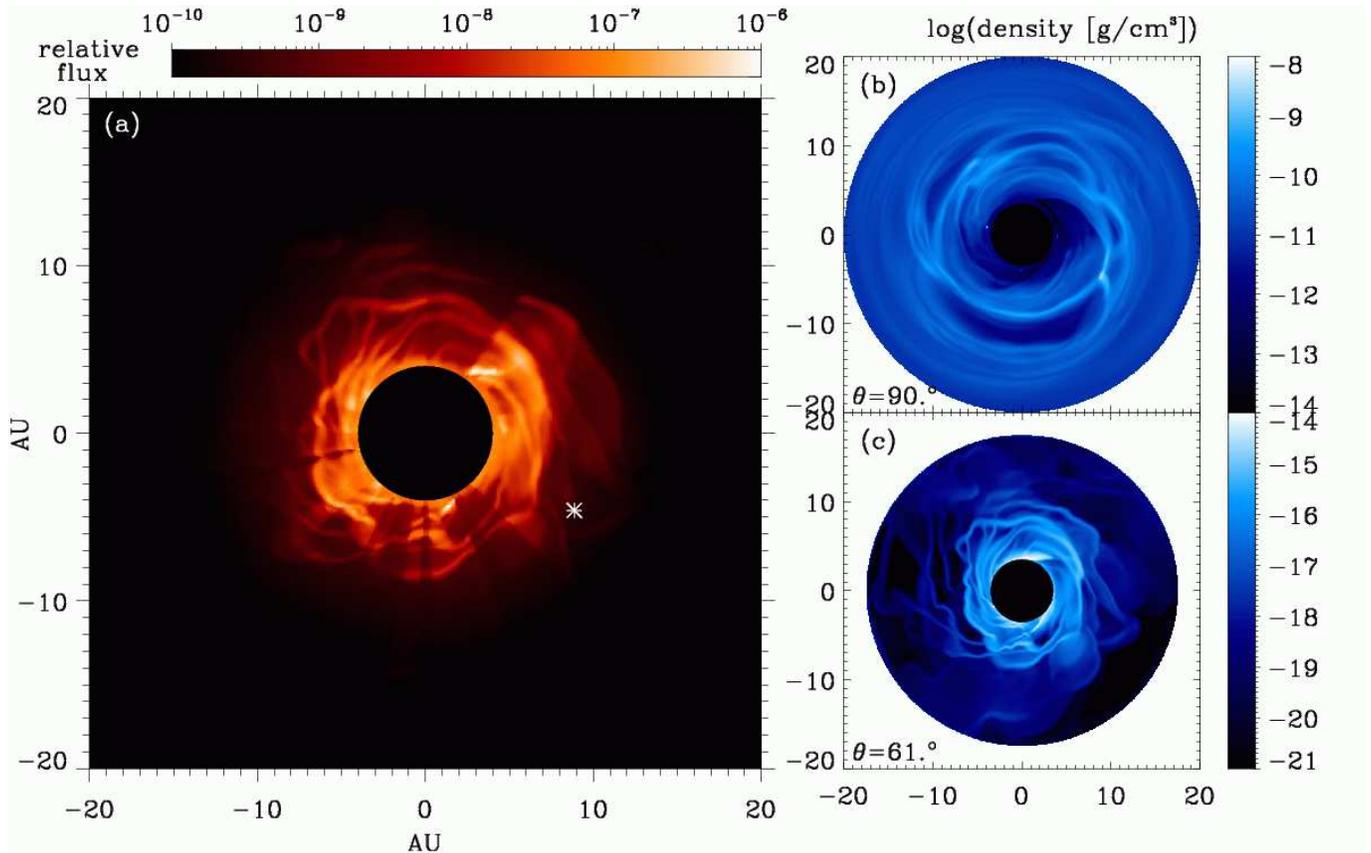}}
\caption{\label{dencompare}
(a) Simulated scattered light image of a planet-forming disk,
normalized to the intensity of stellar emission.  
As indicated by the scale bar, the intensity is displayed in 
the logarithm, with a maximum and minimum of $10^{-6}$ and 
$10^{-10}$, respectively, of the stellar emission.  
The white asterisk shows the location of the forming protoplanet.  
The center 4 AU is outside the simulation region, so it is 
blacked out.
(b) Density at the midplane of the disk.  The forming planet 
can be clearly seen as a high density clump. 
(c) Projected density at an opening angle of $\theta=61\degr$.  
}
\end{figure*}

In \figref{dencompare}a, we show the simulated scattered light image 
of the disk as calculated above, at a time of 339 years.  
The flux is normalized to the 
stellar emission, and is displayed in the logarithm as indicated by the 
color bar.  Note that the brightest intensity
is at best $10^{-6}$ of the stellar emission, 
and the minimum intensity shown is $10^{-10}$ of the stellar emission. 
The dark radial streaks seen in the simulated image are shadows of 
dense clumps at the inner boundary of the simulation, at 4 AU.  
At the distances and scales involved in the simulation, it is 
appropriate to treat the star as a point source, which is why the 
shadows diverge.  
The area interior to 
4 AU has been blacked out to ignore numerical artifacts at the 
inner boundary of the simulation. 
The planet-forming clump is located at the position of the white 
asterisk.  There is no apparent correlation between the position 
of the forming planet and structure in the scattered light image.  

In \figref{dencompare}b, we show the density structure at the disk 
midplane of the simulation.  The density enhancement corresponding to the 
position of the forming planet can be clearly seen.  However, 
the density structure at the disk's midplane bears no resemblance 
to the scattered light image.  
In \figref{dencompare}c, we show the density structure of the 
disk at $\theta=61\degr$, in projection.  
This is close to the height of the scattering surface of the disk.
Comparing figures \ref{dencompare}a and \ref{dencompare}b, it is 
readily apparent that the scattered light emission traces 
the density structure in these tenuous surface layers of the 
disk rather than probing the overall density structure in the disk.  
The densities displayed in figures \ref{dencompare}b and \ref{dencompare}c 
cover completely different ranges, demonstrating this fact.  

The structures seen in scattered light are corrugations 
created in the initially smooth surfaces of the disk from 
the highly dynamical evolution of a marginally
gravitationally unstable disk \markcite{boss01}({Boss} 2001).
The corrugations indicate activity within the disk, 
even though they are not directly correlated spatially.  

In \figref{psfconv}, we demonstrate the time evolution of 
the scattered light image.  From top to bottom, we show 
simulated images of the disk at 335, 339, and 346 years.  
For scale, the central blackout region has a diameter of 8 AU.
The intensity scale is the same as in \figref{dencompare}a and
can be inferred from the contours, which indicate 
flux relative to the stellar emission of 
$10^{-7}$ (solid), $10^{-8}$ (dashed), and $10^{-9}$ (dotted).  
The left column shows the theoretical 
scattered light image, with perfect resolution, with 
the location of the forming planet indicated by the white asterisk.  
These simulated images show that 
the corrugations evolve on time scales of only a few years,
short enough to be easily observationally detectable. Such surface
irregularities are seen in other disk instability simulations as well
\markcite{2006BoleyDurisen}(e.g., {Boley} \& {Durisen} 2006) and can be attributed to spiral shock
fronts breaking on the disk surface.

\section{Discussion: Observability}

\begin{figure*}
\resizebox{\textwidth}{!}{\includegraphics{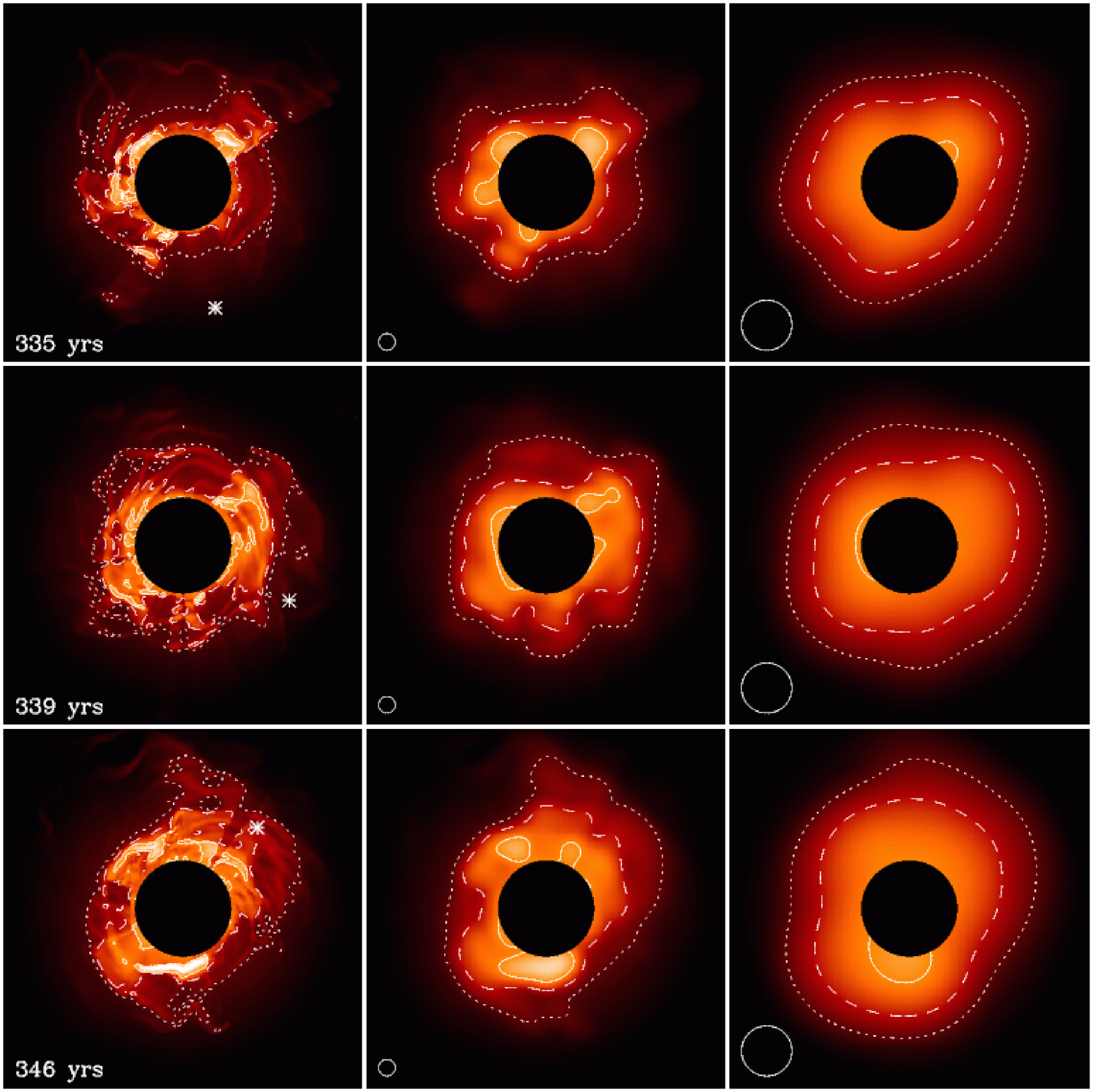}}
\caption{\label{psfconv}Simulated images of a disk undergoing 
disk instablity at a simulation time of 335 (top), 339 (middle) 
and 346 (bottom) years.  The intensity scale is the same 
as in \figref{dencompare}.  
The contours represent flux relative to stellar emission of 
$10^{-7}$ (solid), $10^{-8}$ (dashed), and $10^{-9}$ (dotted).  
The left column shows the theoretical 
emission, with the location of the forming planet indicated by a 
white asterisk.  The middle and right columns show the images when 
convolved with a PSF of $0.01\arcsec$ and $0.03\arcsec$, respectively.  
The white circles represent the sizes of these PSFs.  
We assume that the disk is face-on to the observer 
at a distance of 140 pc. 
}
\end{figure*}

Observations of features indicative of disk instability 
would be a major advance for planet formation theory.  
An 8-m class telescope with adaptive optics would be diffraction 
limited at $0.03\arcsec$ at 1 $\mu$m.  
A 30-m class telescope such as the GMT (Giant Magellan Telescope)
or TMT (Thirty Meter Telescope), which are expected to have first light 
in the next decade, would have diffraction limits of less 
than $0.01\arcsec$ at 1 $\mu$m.  
To illustrate what actual observations of our theoretical disks 
might look like, we have convolved 
the images in the left column of \figref{psfconv} 
with Gaussian PSFs (point spread functions) with 
FWHM (full width at half maximum) of 
$0.01\arcsec$ (center column) and $0.03\arcsec$ (right column). 
We assume that the disk is face-on to 
the observer at 140 pc, the approximate distance of Taurus.

A resolution of $0.01\arcsec$ is just sufficient to begin resolving 
the corrugations in the scattered light image, while 
a resolution of $0.03\arcsec$ reveals only some indistinct hints of structure.  
The Near-Infrared Coronagraphic Imager (NICI) adaptive optics coronograph 
at the Gemini 
Observatory\footnote{http://www.gemini.edu/sciops/instruments/nici/niciIndex.html}
should be diffraction-limited at 1 micron with an 8-m aperture.  
However, its estimated performance is a contrast ratio of 13 mags, or 
$10^{-5.2}$ at $1\arcsec$ from the star, with decreasing performance 
closer to the star.  

This suggests that the spatial features we have 
modeled may not be observable with current 8-m class telescopes, and 
would have to wait until a 30-m class telescope such as the GMT 
comes online.  High-contrast imaging on the GMT, which is being developed 
for imaging exoplanets, should be able to achieve a contrast ratio 
of $10^{-8}$ at $0.05\arcsec$ at 1.65$\mu$m \markcite{2006Angel_etal}({Angel} {et~al.} 2006).  
In comparison, the inner edge of the simulated disk at 4 AU would be 
at $0.06\arcsec$ from the central star.  
This dashed contours in \figref{psfconv} indicate a contrast ratio of 
$10^{-8}$, so the variability in the scattered light patterns 
from disk instability should be observable by the GMT.  
Optical or infrared interferometry might also be good ways to 
observe disk instability.  This method has already been 
used to detect structure in the inner 1 AU of some T Tauri disks 
\markcite{2005ApJ...622..440A}(e.g.~ {Akeson} {et~al.} 2005).

The temporal variations may be observable if 
the disk is revisited over several years.  
Observing this phenomenon requires finding a Class I YSO 
with the right orientation and without too much obscuration by 
an envelope.  Bipolar outflows that are common to many YSOs may 
clear away envelope material along the poles, 
allowing a clear view to the inner regions of the disk that are of 
interest.  

Since the contrast between the disk to the star is $10^{-6}$ at best, 
temporal variations in overall brightness of the unresolved source 
are more likely due to stellar activity than to disk activity.  
If the disk is resolved, then disk activity 
should be readily distinguished from stellar activity 
since the scattered light should scale with stellar brightness.  
Stellar activity that affects the luminosity of the star isotropically 
should not affect the contrast ratio between star and disk.  

Anisotropic variations, such as cold or hot star spots, can be 
distinguished from disk activity because these variations will 
have the same period as the rotation period of the star, which is 
on the order of days for typical young stars.  
For example, HH30 is an edge-on disk system 
whose periodic fluctuations in brightness are caused by 
illumination from a hot spot on the star sweeping across the surface 
of the disk as the star rotates, similar to a lighthouse 
\markcite{1998ApJ...497..404W,1998ApJ...506L..43W,2001ApJ...556..958C}({Wood} {et~al.} 1998; {Wood} \& {Whitney} 1998; {Cotera} {et~al.} 2001).  
This variability occurs on time scales of days and is periodic.  
In contrast, the features we have modeled  
vary over years and are irregular.  
By monitoring an object over a series of several nights, short-term 
effects can be filtered out and the year-to-year variations 
can be studied to determine if activity characteristic of 
disk instability is occuring.  

In contrast to the variability characteristic of 
disk instability, planet formation by core accretion should be 
more quiescent, resulting in a distinct shadow at the position of the 
forming planet core \markcite{2005HJC_AAS,2005HJC_PPV}({Jang-Condell} 2005a, 2005b).  The mechanisms are 
also separated in pre-main sequence evolutionary epochs: 
disk instability is expected to occur during the Class I phase of 
star/planet formation, where there is substantial disk and envelope 
material infalling toward the star.  Core formation has a much longer 
expected time scale, so those signatures are detectable during 
the Class II phase, after the envelope has cleared but the disk 
still remains.

\section{Conclusions} 

Since scattered light images of optically thick disks 
give information about the tenuous upper layers of the 
disk rather than the overall large-scale structure, 
one should be cautious when trying to 
infer disk structure from scattered light images.  
For example, the outer disk of AB Aurigae \markcite{2004Fukagawa_etal}({Fukagawa} {et~al.} 2004) 
extends out to hundreds of AU, and 
over that path length, even a small amount of material can easily 
block stellar illumination.  
This effect should be considered before drawing strong conclusions 
about the underlying disk structure.  

Although scattered light emission is not directly correlated to 
the planet forming in the disk, the presence of 
structure in the scattered light emission coupled with 
high time variability can be an indicator of planet forming activity.  
We can at least identify disks where planets are forming even 
if we cannot say exactly where in the disk it is taking place.
The GMT 
might well be able to detect these rapid variations in scattered
light on spatial scales of interest ($\sim$ 5 to 10 AU) for
giant planet formation, and thus shed light on the
processes occuring deep inside optically thick protoplanetary
disks.

Emission from the disk itself at longer wavelengths should 
be a better probe of the overall disk structure both because 
the optical depth is smaller at longer wavelength and because 
direct emission from the disk has better contrast with 
respect to the star than scattered light.  Observations of 
gravitationally unstable disks in the millimeter and sub-millimeter, 
such as those predicted by \markcite{2006Narayanan}({Narayanan} {et~al.} 2006), will be a useful 
complement to those predicted in this paper.  

\acknowledgements

We thank John Chambers for helpful discussions in the preparation 
of this paper, and our anonymous 
referee for helpful comments that greatly improved it.  
The authors acknowledge support 
by the NASA Astrobiology Institute under
Cooperative Agreement NNA04CC09A.

\bibliography{}

\end{document}